\documentclass[prl,aps,twocolumn,groupedaddress]{revtex4-1}

\usepackage{bm}
\usepackage{graphicx}
\usepackage{color}
\usepackage{ulem}

\begin{document}

\title{Carrier Injection and Manipulation of Charge-Density Wave in Kagome Superconductor CsV$_3$Sb$_5$}

\author{Kosuke Nakayama,$^{1,2,\#,\ast}$ Yongkai Li,$^{3,4,\#}$ Takemi Kato,$^1$ Min Liu,$^{3,4}$ Zhiwei Wang,$^{3,4,\ast}$ Takashi Takahashi,$^{1,5,6}$ Yugui Yao,$^{3,4}$ and Takafumi Sato$^{1,5,6,}$}
\email[\#These authors equally contributed to this work.\\
$^{\ast}$Corresponding authors:]{k.nakayama@arpes.phys.tohoku.ac.jp; zhiweiwang@bit.edu.cn; t-sato@arpes.phys.tohoku.ac.jp}
\affiliation{$^1$Department of Physics, Tohoku University, Sendai 980-8578, Japan\\
$^2$Precursory Research for Embryonic Science and Technology (PRESTO), Japan Science and Technology Agency (JST), Tokyo, 102-0076, Japan\\
$^3$Centre for Quantum Physics, Key Laboratory of Advanced Optoelectronic Quantum Architecture and Measurement (MOE), School of Physics, Beijing Institute of Technology, Beijing 100081, China\\
$^4$Beijing Key Lab of Nanophotonics and Ultrafine Optoelectronic Systems, Beijing Institute of Technology, Beijing 100081, China\\
$^5$Center for Spintronics Research Network, Tohoku University, Sendai 980-8577, Japan\\
$^6$WPI Research Center, Advanced Institute for Materials Research, Tohoku University, Sendai 980-8577, Japan\\
}

\date{\today}

\begin{abstract}
Kagome metals $A$V$_3$Sb$_5$ ($A$ = K, Rb, and Cs) exhibit a unique superconducting ground state coexisting with charge-density wave (CDW), whereas how these characteristics are affected by carrier doping remains unexplored because of the lack of an efficient carrier-doping method. Here we report successful electron doping to CsV$_3$Sb$_5$ by Cs dosing, as visualized by angle-resolved photoemission spectroscopy. We found that the electron doping with Cs dosing proceeds in an orbital-selective way, as characterized by a marked increase in electron filling of the Sb 5$p_z$ and V 3$d_{xz/yz}$ bands as opposed to relatively insensitive nature of the V 3$d_{xy/x^2-y^2}$ bands. By monitoring the temperature evolution of the CDW gap around the $\bar{\rm M}$ point, we found that the CDW can be completely killed by Cs dosing while keeping the saddle point with the V 3$d_{xy/x^2-y^2}$ character almost pinned at the Fermi level. The present result suggests a crucial role of multi-orbital effect to the occurrence of CDW, and provides an important step toward manipulating the CDW and superconductivity in $A$V$_3$Sb$_5$.
\end{abstract}

\pacs{}

\maketitle
Kagome lattice is an excellent playground to study the physics intertwining electron correlation and topology, owing to its peculiar band structure. Single-orbital model calculations with the kagome lattice predict the formation of nearly flat band, Dirac-cone band, and saddle-point van Hove singularity. When either of these bands is tuned near the Fermi level ($E_{\rm F}$), various exotic states would be realized, such as Weyl magnet \cite{NayakSA2016, YangNJP2017, KurodaNM2017, YeNature2018, LiuNP2018, MoraliScience2019, LiuScience2019}, density wave orders \cite{IsakovPRL2006, GuoPRB2009, WangPRB2013}, charge fractionalization \cite{OBrienPRB2010, RueggPRB2011}, and superconductivity \cite{WangPRB2013, KoPRB2009, KieselPRB2012, KieselPRL2013}. These intriguing characteristics that depend on the electron band filling make the kagome lattice an excellent platform to explore the novel quantum states.

Recent discovery of superconductivity with the superconducting transition temperature $T_{\rm c}$ of 0.93-2.5 K in a family of kagome metals $A$V$_3$Sb$_5$ (AVS: $A$ = K, Rb, and Cs) \cite{OrtizPRM2019, OrtizPRL2020, OrtizPRM2021} sparked a great deal of attention, because the realization of superconductivity in kagome metals is very rare and may invoke unconventional mechanism as inferred from possible symmetry breaking \cite{JiangAX2020, HChenAX2021, HZhaoAX2021, XiangAX2021}. Additionally, AVS exhibits a charge-density wave (CDW) transition at $T_{\rm CDW}$ = 78-103 K accompanied with three-dimensional (3D) 2$\times$2$\times$2 charge order \cite{LiAX2021, LiangAX2021}. From angle-resolved photoemission spectroscopy (ARPES) and density-functional-theory (DFT) calculations \cite{OrtizPRL2020, LiAX2021, LiuAX2021, YangSA2020, WangAX2021, NakayamaAX2021}, the band structure of AVS was found to be characterized by the existence of multi-orbital bands apart from the simple single-orbital model, (i) kagome-lattice bands of the V 3$d$ orbitals forming saddle points near $E_{\rm F}$ at the $\bar{\rm M}$ point and Dirac-cone bands, and (ii) an Sb 5$p$ band forming an electron pocket at the  $\bar{\rm \Gamma}$ point of the surface Brillouin zone.

A central issue under intensive debate is the relationship between the band structure and the mechanism of CDW and superconductivity. Theoretically, the saddle point can promote $f$- or $d$-wave superconducting pairing associated with the scattering with $Q$ = ($\pi$, 0) vector connecting the saddle points, whereas this scattering would also enhance an instability toward CDW (possibly chiral CDW) or unconventional density waves \cite{KieselPRB2012, KieselPRL2013, NandkishoreNP2012, TanAX2021, FengAX2021, LinAX2021, WuAX2021}. Thus, the interplay between CDW and superconductivity is not straightforward and a fundamental question as to whether the saddle points play a more important role in CDW or superconductivity needs to be answered. Given the multi-orbital character of AVS, it is also essential to understand how other bands/orbitals are involved in CDW and superconductivity. In the experimental side, the observation of a CDW gap \cite{JiangAX2020, HChenAX2021, HZhaoAX2021, NakayamaAX2021, ZhouAX2021}, particularly the large gap on the saddle point of the V $d_{xy/x^2-y^2}$ band \cite{NakayamaAX2021, ZhouAX2021}, supports that the saddle points play a certain role in stabilizing CDW. However, the mechanism of CDW and superconductivity is still far from reaching a consensus. For example, the influence of multi-orbital character is yet to be clarified. A promising route to clarify this issue is to control the band energy position relative to $E_{\rm F}$ through carrier doping because the CDW transition would be sensitive to the saddle-point energy and/or the band filling in the simple nesting picture. Carrier doping may also modify a charge balance among multiple orbitals and would be useful to differentiate the contribution of each orbital, as has been utilized to pin down the origin of multi-orbital superconductivity and density wave in Fe-based superconductors \cite{RichardRPP2011}. However, the carrier-doping effect has been unexplored in AVS due to the lack of an effective means of carrier doping.

In this article, we report a high-resolution ARPES study on Cs-dosed CsV$_3$Sb$_5$ (CVS) bulk single crystals. We have succeeded in doping electron carriers via surface decoration, i.e. Cs dosing, and dramatically altering the CDW properties. By experimentally establishing the evolution of Fermi surface (FS) and band structure upon electron doping, we uncovered the orbital-dependent electron-doping effect and suppression of CDW accompanied with the reduction of CDW gaps. We discuss implications of the present results in relation to the mechanism of CDW and superconductivity in AVS.

High-quality single crystals of CVS were synthesized with the self-flux method \cite{OrtizPRM2019}. ARPES measurements were performed using Scienta-Omicron SES2002 spectrometer with the He discharge lamp ($h\nu$ = 21.218 eV) at Tohoku University and Scienta-Omicron DA30 spectrometer with energy tunable photons ($h\nu$ = 85-125 eV) at BL-28A in Photon Factory, KEK. The energy/angular resolution was set to be 7-30 meV/0.2$^{\circ}$ at Tohoku University and 35-60 meV/0.3$^{\circ}$ at Photon Factory. $E_{\rm F}$ of the sample was referenced to that of a gold film evaporated onto the sample holder. We cleaved the sample in ultrahigh vacuum of $<1\times10^{-10}$ Torr and immediately carried out the Cs dosing by using Cs dispenser (SAES Getters) at room temperature, and then cooled down the sample to low temperature for ARPES measurements. It is noted that, ideally, half of the cleaved surface is terminated by Cs atoms in CVS. After Cs dosing, ARPES likely probes only the Cs-terminated surface. Also, the star-of-David or inverse star-of-David distortion would not disarrange the position of Cs atoms, so that Cs atoms evaporated on the CVS surface likely form the 1$\times$1 structure without formation of superlattice. This expectation is consistent with the absence of band folding due to the superstructure in Cs-dosed samples as shown later. The first-principles band-structure calculations were carried out using the full-potential linearized augmented plane-wave method implemented in the WIEN2K code \cite{Blaha2013} with generalized gradient approximation (GGA) \cite{PerdewPRL1996} and the Perdew-Burke-Ernzerhof (PBE) \cite{PerdewPRL1997} type exchange-correlation potential. Lattice constants were fully optimized. Spin-orbit coupling (SOC) was neglected to discuss the orbital-dependent band shift without a complication by the SOC-induced hybridization.

\begin{figure}
\includegraphics[width=3.4in]{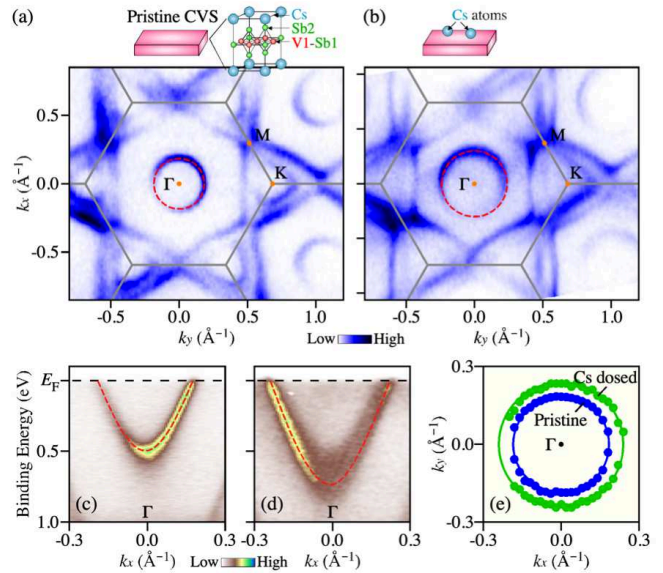}
\vspace{0cm}
\caption{(Color online) (a), (b) ARPES-intensity map at $E_{\rm F}$ plotted as a function of $k_x$ and $k_y$ measured at $T$ = 120 K (above $T_{\rm CDW}$) with 106-eV photons (corresponding to the $k_z$ $\sim$ 0 plane) for pristine CVS (pristine) and moderately Cs-dosed (Cs) samples, respectively. Inset to (a) shows the crystal structure of CVS. (c), (d) Comparison of the ARPES intensity around the $\Gamma$ point between pristine and Cs samples. (e) Comparison of the {\bf k}$_{\rm F}$ points for the $\Gamma$-centered pocket between the two samples.}
\end{figure}

\begin{figure*}
\includegraphics[width=6.5in]{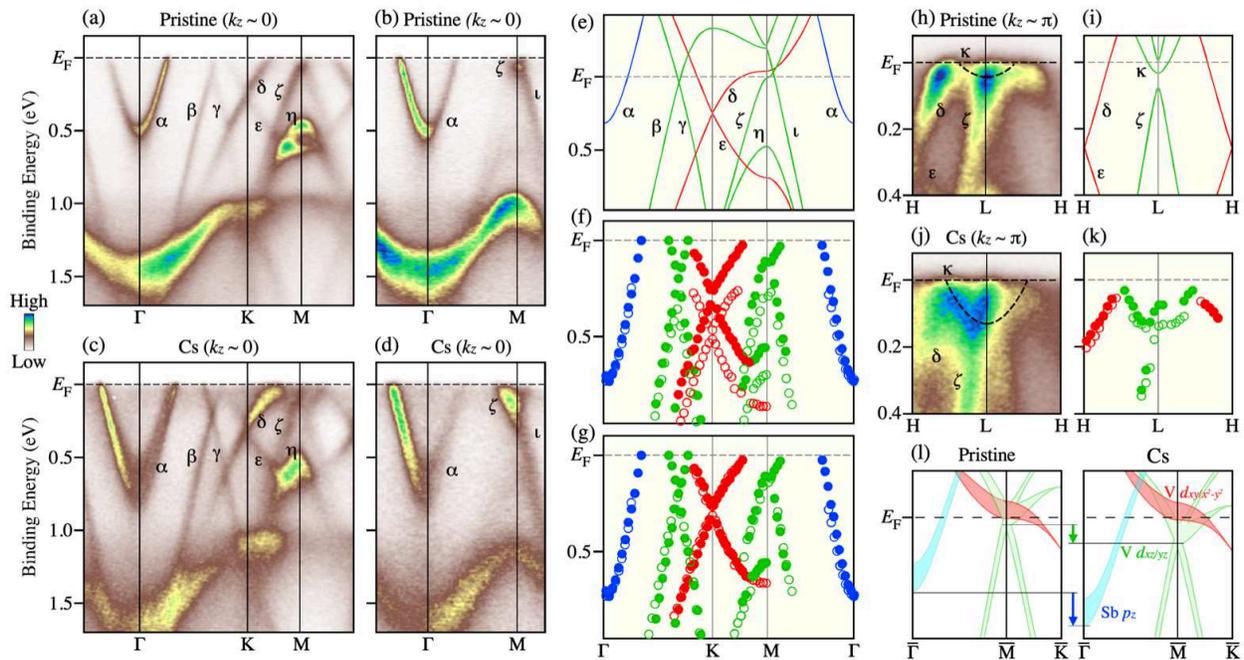}
\vspace{0cm}
\caption{(Color online) (a), (b) ARPES intensity as a function of wave vector and binding energy ($E_{\rm B}$) in a wide $E_{\rm B}$ range, measured at $T$ = 120 K along the $\Gamma$KM and $\Gamma$M lines, respectively, for pristine sample. (c), (d) Same as (a) and (b), respectively, but for Cs sample. (e) The calculated band structure along the $\Gamma$KM cut. Dominant orbital character of each band is indicated by coloring (blue, Sb $p_z$; green, V $d_{xz/yz}$; red, V $d_{xy/x^2-y^2}$). Small hybridization at each band intersection was removed from the calculated bands to highlight orbital character of each band. Calculated $E_{\rm F}$ is shifted downward by 90 meV to obtain a better matching with the experimental data. (f) Comparison of experimental band dispersions between pristine and Cs samples (open and filled circles, respectively) extracted from the peak position in the ARPES spectra of (a)-(d). The band dispersion for pristine sample is shifted downward as a whole by 240 meV. (g) Same as (f) but the band dispersion of pristine sample incorporates the band-dependent energy shift; 240 meV for the Sb $p_z$ band (blue), 100 meV for the V $d_{xz/yz}$ band (green), and 40 meV for the V $d_{xy/x^2-y^2}$ band (red). (h) ARPES intensity at $T$ = 120 K for pristine sample measured along the HL line with $h\nu$ = 120 eV. (i) Calculated band structure in the normal state along the HL cut. Calculated $E_{\rm F}$ is shifted downward by 90 meV to obtain a better matching with the experimental data. (j) Same as (h) but after Cs dosing. (k) Same as (g) but for the HL cut. (j) Schematics of band-dependent energy shifts with Cs dosing. $k_z$ dispersion of the bulk bands is shown by shaded area.}
\end{figure*}

We at first present the evolution of FS with Cs deposition. Figures 1(a) and 1(b) show the FS mapping for pristine and moderately Cs-dosed (labelled as Cs) samples obtained at $T$ = 120 K above $T_{\rm CDW}$ of pristine CVS (91 K) with 106-eV photons which probe the electronic states at $k_z$ $\sim$ 0 \cite{NakayamaAX2021}. In pristine sample, a circular pocket formed by a parabolic electron band centered at the $\Gamma$ point [Fig. 1(d)] is recognized. This pocket is attributed to the 5$p_z$ band of Sb1 atoms embedded in the kagome-lattice plane [inset to Fig. 1(a)] \cite{LiuAX2021, JiangAX2020, JZhaoAX2021}. At the Brillouin-zone boundary, two features are recognized, one forms a large hexagonal FS centered at the $\Gamma$ point with mainly V 3$d_{xz/yz}$ character of the kagome-lattice band and the other forms a triangular FS centered at the K point with mainly V 3$d_{xy/x^2-y^2}$ character \cite{LiuAX2021, JZhaoAX2021}. A side-by-side comparison of Figs. 1(a) and 1(b) reveals that the $\Gamma$-centered pocket expands with Cs doping (dashed circles). This is natural because each Cs atom adsorbed on the surface would donate one electron to the CVS bands. The FS expansion is also visualized by a direct comparison of the parabolic band around the $\Gamma$ point in Figs. 1(c) and 1(d) showing a clear downward shift with Cs dosing. By a quantitative analysis of the experimental band dispersions, the energy shift of Cs sample with respect to the pristine one was estimated to be $\sim$0.24 eV, indicating a marked influence of Cs dosing to the surface electronic states of CVS. By determining the location of Fermi-wave-vector ({\bf k}$_{\rm F}$) points in the 2D {\bf k} space for the electron pocket [Fig. 1(e)], the FS for Cs sample was found to be expanded by 65 \% relative to that of pristine sample.

To examine the energy shift of bands in more detail, we show the APRES intensity along the $\Gamma$KM and $\Gamma$M cuts in pristine and Cs samples at $T$ = 120 K in Figs. 2(a)-2(d). The ARPES intensity for pristine sample in Figs. 2(a) and 2(b) signifies the direct correspondence with the calculated bands shown in Fig. 2(e) (blue, green, and red coloring indicates the dominant contribution of the Sb $p_z$, V $d_{xz/yz}$, and V $d_{xy/x^2-y^2}$ orbital, respectively). For example, besides the $\Gamma$-centered electron band with the Sb $p_z$ orbital (labelled as $\alpha$ band) seen in Fig. 1(c), two linearly dispersive bands ($\beta$ and $\gamma$ bands) with the V $d_{xz/yz}$ character forming a Dirac-cone-like dispersion between the $\Gamma$ and K points are identified in the experiment [Fig. 2(a)]. Also, the calculated $d_{xy/x^2-y^2}$ band ($\delta$ band) forming a saddle point pinned almost at $E_{\rm F}$, called here the saddle-point band, is seen in the experiment as a shallow hole band at the M point. Another calculated $d_{xy/x^2-y^2}$ band ($\epsilon$ band) that intersects the saddle-point band at the K point to form a Dirac point (in the case of negligible SOC) is also seen in the experiment. The ARPES intensity also reproduces the calculated $d_{xz/yz}$ bands topped at $\sim$0.45 eV and near $E_{\rm F}$ at the M point ($\zeta$ and $\eta$ bands) along the KM cut [Fig. 2(a)], and a shallow electron band ($\zeta$ band) and a holelike band with the top of dispersion slightly below $E_{\rm F}$ ($\iota$ band) at the M point along the $\Gamma$M cut [Fig. 2(b)]. These experimental results show a good agreement with the calculation regarding the number and shape of the band dispersion. All the near-$E_{\rm F}$ bands predicted in the calculation are also resolved in Cs sample [Figs. 2(c) and 2(d)] but with a band-dependent energy shift compared with pristine sample. For instance, the crossing point between the $\beta$ and $\gamma$ bands shifts downward by $\sim$0.1 eV after Cs dosing, in disagreement with the downward shift of 0.24 eV for the $\Gamma$-centered $\alpha$ band. Furthermore, the crossing point between the $\delta$ and $\epsilon$ bands does not show a clear downward shift by Cs dosing [Fig. 2(c)], suggesting the orbital-dependent band shift.

\begin{figure}
\includegraphics[width=3.4in]{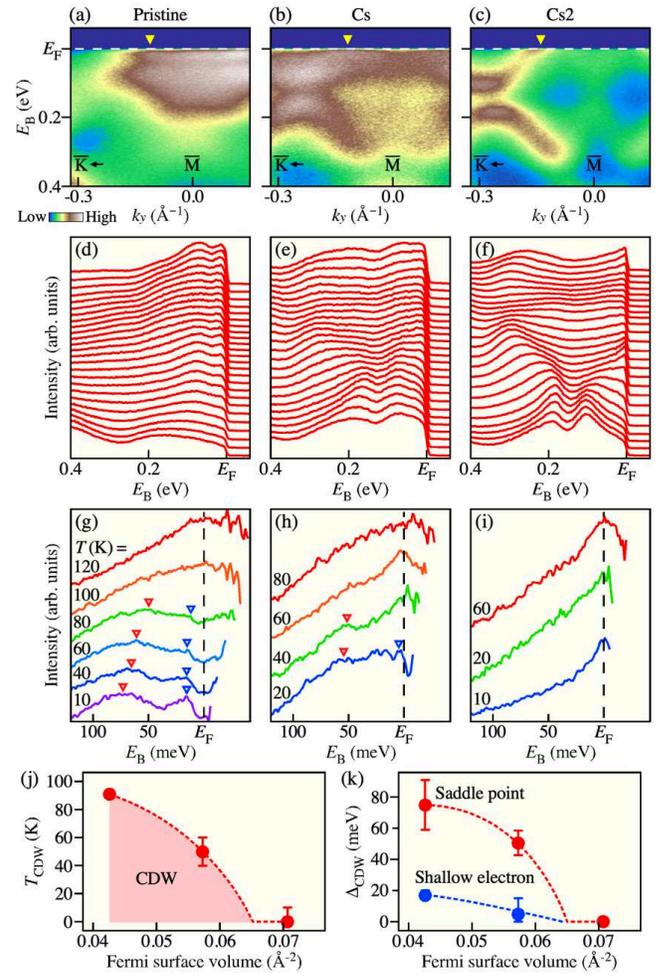}
\vspace{0cm}
\caption{(Color online) (a)-(c) ARPES intensity and (d)-(f) corresponding EDCs, measured along the $\bar{\rm M}\bar{\rm K}$ cut at $T$ = 10-20 K for three samples. (g)-(i) Temperature dependence of the EDCs divided by the FD function convoluted with an instrumental resolution for the three samples. (j), (k) Experimental $T_{\rm CDW}$ and CDW-gap size $\Delta$, respectively, plotted against the volume of electron pocket at $\bar{\Gamma}$. $\Delta$ was determined by tracing the peak position of EDCs at 10 K for pristine and Cs2 samples and at 20 K for Cs sample, and plotted for both the saddle-point and shallow electron bands.}
\end{figure}

The orbital-dependent band shift is more clearly seen in the direct comparison of the experimental band dispersions. When we compare the band structure of Cs sample [filled circles in Fig. 2(f)] with that of pristine sample shifted by 240 meV in a rigid-band manner (open circles), the $\alpha$ band with the Sb $p_z$ character shows a good agreement between the two (see blue circles), but the V 3$d$ bands show a fatal mismatch in the energy position (green and red circles). On the other hand, as highlighted in Fig. 2(g), when we shift the V $d_{xz/yz}$ bands by 100 meV and the $d_{xy/x^2-y^2}$ bands by 40 meV, the band structure shows an excellent matching between pristine and Cs samples. Thus, the present ARPES measurements unambiguously support the orbital-dependent band shift. The existence of the band-dependent energy shift is corroborated by the same argument for the $k_z$ $\sim$ $\pi$ plane, e.g., along the HL cut [Figs. 2(h)-2(k)], a shallow electron pocket ($\kappa$ band) appears in the vicinity of $E_{\rm F}$ [see black dashed curve in Fig. 2(h)] in addition to the $\delta$, $\epsilon$, and $\zeta$ bands, and their energy positions after Cs dosing [Fig. 2(j)] match well with those of pristine sample by incorporating the orbital-dependent band shift of 240 meV and 100 meV for the Sb $p_z$ and V $d_{xz/yz}$ bands, respectively [Fig. 2(k)]. It is emphasized here that even when the $p_z$ and $d_{xz/yz}$ bands including the shallow electron pocket at the L point are heavily electron doped after Cs dosing, the saddle point at the M point is still located at around $E_{\rm F}$ owing to the much smaller energy shift of the $d_{xy/x^2-y^2}$ band, as schematically shown in Fig. 2(l); specifically, the $d_{xy/x^2-y^2}$-derived saddle point is still located slightly above $E_{\rm F}$ after Cs dosing (see Supplementary Material \cite{SM}). This observation has an important implication to the mechanism of CDW, as discussed later.

Now that the evolution of band structure with Cs dosing is established in the normal state, the next important question is whether or not the CDW properties are affected by the Cs dosing. For this sake, we cooled the sample down to well below $T_{\rm CDW}$ of pristine sample (91 K), and investigated possible existence of the CDW gap by high-energy-resolution measurements with lower-energy photons ($h\nu$ = 21.2 eV). Figures 3(a)-3(c) show a comparison of the ARPES intensity along the $\bar{\rm M}$$\bar{\rm K}$ cut at $T$ = 10-20 K among pristine, moderately Cs-dosed (Cs), and heavily Cs-dosed (Cs2) samples (note that moderately Cs-dosed sample shown in Fig. 3 is a different sample from Cs sample in Figs. 1 and 2, but the doping level is almost identical, so we call it Cs sample). Corresponding EDCs are shown in Figs. 3(d)-3(f). In pristine sample, there exists a characteristic peak-dip-hump structure in the EDC in Fig. 3(d) and a dip-originated weaker intensity region at $\sim$35 meV in the ARPES intensity in Fig. 3(a). The peak-dip-hump structure is associated with the formation of CDW \cite{NakayamaAX2021} because it appears only below $T_{\rm CDW}$ as shown by the temperature dependence of EDCs measured at the {\bf k}$_{\rm F}$ point of the shallow electron band [yellow triangle in Fig. 3(a)] in Fig. 3(g), in which the EDCs have been divided by the Fermi-Dirac distribution (FD) function to eliminate the influence from the cut-off by the FD function. The EDC at $T$ = 10 K consists of two distinct peaks, a hump at $\sim$70 meV and a sharp peak at $\sim$20 meV, which correspond to the CDW gaps for the saddle-point band ($k_z$ $\sim$ 0) and the shallow electron band ($k_z$ $\sim$ $\pi$), respectively, as discussed in our previous study \cite{NakayamaAX2021} (also see Supplemental Material for more information on the validity of our assignment \cite{SM}).

In Cs sample, a peak-dip-hump structure is still recognized at $T$ = 20 K [Fig. 3(e)] although it is less pronounced compared to the case of pristine sample. As shown in Fig. 3(h), the peak-dip-hump structure is seen at 20 and 40 K, but it disappears at 60 K. This suggests that $T_{\rm CDW}$ for Cs sample is between 40 and 60 K, which is reduced by $\sim$35 \% compared to that of pristine sample (note that $T_{\rm CDW}$ refers to the $T_{\rm CDW}$ at the surface layer since $T_{\rm CDW}$ for the bulk would be unaffected by the Cs dosing). The EDC in Fig. 3(h) also signifies that the CDW gap for the saddle-point band is $\sim$50 meV. A signature of small but finite gap opening may be found for the shallow electron band at 20 K, judging from a clear suppression of the spectral weight at $E_{\rm F}$ compared with that at high temperatures. This indicates that the CDW gaps for both the saddle-point and shallow electron bands are reduced after Cs deposition, consistent with the reduction of $T_{\rm CDW}$.

We found that the CDW can be completely suppressed by further Cs dosing. As shown in Figs. 3(c) and 3(f), there exists no anomaly at $T$ = 10 K in the ARPES intensity, and the peak-dip-hump structure is absent in the EDC for Cs2 sample [note that the absence of the peak-dip-hump structure in Cs2 sample is unlikely due to the the spectral broadening by the disorder effect because Cs2 sample shows a sharper quasiparticle peak than the pristine sample, as recognized from a comparison between Figs. 3(g) and 3(i); for details, see Supplemental Material \cite{SM}]. The absence of CDW-related feature is also confirmed by the temperature dependence of EDCs in Fig. 3(i) (see Supplemental Material for the absence of a CDW gap irrespective of \textbf{k} position \cite{SM}). We summarize in Fig. 3(j) the $T_{\rm CDW}$ estimated from the EDCs in Figs. 3(g)-3(i) and the CDW-gap size $\Delta$ for both the saddle-point and shallow electron bands, plotted as a function of the volume of electron pocket at $\Gamma$. Obviously, the electron doping leads to the simultaneous reduction of $T_{\rm CDW}$ and $\Delta$, indicating that the CDW is sensitive to the carrier concentration.

Now we discuss implications of the present results in relation to CDW and superconductivity. We found from Figs. 2 and 3 that the CDW can be completely suppressed by Cs dosing while pinning the saddle point with the $d_{xy/x^2-y^2}$ orbital almost at $E_{\rm F}$. In contrast to the almost stationary behavior of the saddle point, the electron band with the $d_{xz/yz}$ orbital is shifted downward away from $E_{\rm F}$ by Cs dosing. These observations indicate that, although the large gap opening on the saddle point plays a certain role in lowering the electronic energy to stabilize CDW, the suppression of CDW by Cs dosing cannot be explained by a simple scenario that moving the van-Hove singularity away from $E_{\rm F}$ kills CDW. Our results reveal that the stability of CDW is governed by the energy position of not only the saddle-point band but also the $d_{xz/yz}$-derived band. This would be reasonable because the electron band with the $d_{xz/yz}$ orbital is connected to itself by $Q$ = ($\pi$, 0) when the band bottom is shallow, thereby participating in the CDW formation. In addition, existence of the scattering channel between the saddle-point and shallow electron bands at the M and L points, respectively, may be linked to the 3D nature of CDW. Our result suggests that, besides the $d_{xy/x^2-y^2}$ orbital which shows the largest CDW gap, the $d_{xz/yz}$ orbitals with a smaller gap are also of crucial importance for the occurrence of CDW, pointing to the importance of multi-orbital effect. Therefore, although most of previous theoretical models on the mechanisms of CDW and superconductivity in AVS were constructed with a single orbital, either $d_{xy/x^2-y^2}$ or $d_{xz/yz}$, the contribution from both $d_{xy/x^2-y^2}$ and $d_{xz/yz}$ orbitals should be considered to construct a microscopic theory to explain the exotic properties of AVS. Since a key parameter that controls the multi-orbital effect is thought to be an energy/carrier balance between different orbitals, we have controlled such a balance by the Cs dosing onto the crystal surface. The observed orbital-dependent band shift can be understood in terms of the difference in the spatial distribution of each orbital; namely, the orbital extending toward out of kagome plane shows the larger energy shift compared to the orbital extending within the kagome plane, as suggested from the large shift for the Sb $p_z$, the moderate shift for the V $d_{xz/yz}$ orbital, and the small shift for the V $d_{xy/x^2-y^2}$ orbital. This would be natural if electrons provided by Cs atoms accumulate near the surface. Since a similar inhomogeneous carrier distribution along the $z$ axis is expected for thin films/flakes under the electrical gating, experiments with gating on AVS would be useful to study the multi-orbital physics. The multi-orbital effect may be also important to understand the alkaline-metal dependence of physical properties in AVS, because the relative energy position between $d_{xz/yz}$ and $d_{xy/x^2-y^2}$ orbitals seems to be alkali-metal dependent according to the band-structure calculations. Besides the importance of the multi-orbital effect, the largely enhanced density of states at $E_{\rm F}$ due to the proximity of the saddle point to $E_{\rm F}$ without the disturbance from the likely-competing CDW (that suppresses the density of states at $E_{\rm F}$) revealed in Cs2 sample suggests an intriguing possibility that the carrier tuning via alkaline-metal dosing (or the electrical gating for thin films/flakes and chemical replacement for the bulk crystals) serves as an effective means to manipulate not only CDW but also superconductivity, hopefully leading to an enhancement of $T_{\rm c}$. Clarifying the validity of this argument awaits further experiments for carrier-modulated AVS.

In conclusion, the present ARPES study has established that Cs dosing serves as a useful means to control the electron band filling and the CDW properties in CsV$_3$Sb$_5$. We revealed that Cs dosing leads to an intriguing change in the band structure that is beyond the explanation with the rigid-band model, i.e. the orbital-selective electron doping. We found that the CDW can be completely killed when the $d_{xz/yz}$-derived shallow electron band is pushed away from $E_{\rm F}$ while keeping the saddle point of the $d_{xy/x^2-y^2}$ orbital pined almost at $E_{\rm F}$, pointing to the importance of the multi-orbital effect for understanding the CDW. The present result opens a pathway toward manipulating the CDW and superconducting properties in $A$V$_3$Sb$_5$ through carrier tuning.

\begin{acknowledgments}
This work was supported by JST-CREST (No. JPMJCR18T1), JST-PRESTO (No. JPMJPR18L7), and Grant-in-Aid for Scientific Research (JSPS KAKENHI Grant Numbers JP17H01139, JP18H01160). The work at Beijing was supported by the National Key R\&D Program of China (Grant No. 2020YFA0308800), the Natural Science Foundation of China (Grant Numbers 92065109, 11734003, 12061131002), the Beijing Natural Science Foundation (Grant No. Z190006), and the Beijing Institute of Technology (BIT) Research Fund Program for Young Scholars (Grant No. 3180012222011). T. K. acknowledges support from GP-Spin. Z.W thanks the Analysis \& Testing Center at BIT for assistances in facility support. 
\end{acknowledgments}

\bibliographystyle{prsty}

\end{document}